\documentclass[journal]{IEEEtran}

\ifCLASSINFOpdf

\else

\fi

\usepackage{cite}
\usepackage{amsmath,amssymb,amsfonts}
\usepackage{graphicx}
\usepackage{textcomp}
\usepackage[english]{babel}
\usepackage{blindtext}
\usepackage[utf8]{inputenc}
\usepackage{xcolor}
\usepackage{colortbl}
\usepackage[linesnumbered,algo2e,ruled]{algorithm2e}
\usepackage[inner=2.5cm,outer=2.5cm,top=3cm,bottom=3cm]{geometry}
\usepackage{tabularx}
\usepackage[thinlines]{easytable}
\usepackage{tablefootnote}
\usepackage{longtable}
\usepackage{multirow}
\usepackage[hyphens]{url}
\usepackage[hidelinks, bookmarks=false]{hyperref}
\usepackage{siunitx}
\usepackage{amsthm}
\usepackage{booktabs}
\usepackage{dcolumn}
\usepackage{subcaption}
\usepackage{mathtools, nccmath}
\usepackage{tikz}
\newcommand{\ballnumber}[1]{\tikz[baseline=(myanchor.base)] \node[circle,fill=.,inner sep=1pt] (myanchor) {\color{-.}\bfseries\footnotesize #1};}
\DeclarePairedDelimiter{\nint}\lfloor\rceil

\newcommand\copyrighttext{%
  \footnotesize \textcopyright 2019 IEEE. Personal use of this material is permitted.
  Permission from IEEE must be obtained for all other uses, in any current or future
  media, including reprinting/republishing this material for advertising or promotional
  purposes, creating new collective works, for resale or redistribution to servers or
  lists, or reuse of any copyrighted component of this work in other works.}
\newcommand\copyrightnotice{%
\begin{tikzpicture}[remember picture,overlay]
\node[anchor=south,yshift=10pt] at (current page.south) {\fbox{\parbox{\dimexpr\textwidth-\fboxsep-\fboxrule\relax}{\copyrighttext}}};
\end{tikzpicture}%
}

\begin{document}

\title{Hardening Random Forest Cyber Detectors Against Adversarial Attacks}

\author{\IEEEauthorblockN{Giovanni Apruzzese, Mauro Andreolini, Michele Colajanni, Mirco Marchetti}\\
\IEEEauthorblockA{\textit{Department of Engineering ``Enzo Ferrari''}\\
\textit{University of Modena and Reggio Emilia}\\
Modena, Italy\\
\{giovanni.apruzzese, mauro.andreolini, michele.colajanni, mirco.marchetti\}@unimore.it}}

\markboth{}%
{Shell \MakeLowercase{\textit{et al.}}: Bare Demo of IEEEtran.cls for IEEE Journals}

\maketitle
\copyrightnotice
\begin{abstract}
Machine learning algorithms are effective in several applications, but they are not as much successful when applied to intrusion detection in cyber security. Due to the high sensitivity to their training data, cyber detectors based on machine learning are vulnerable to targeted adversarial attacks that involve the perturbation of initial samples. Existing defenses assume unrealistic scenarios; their results are underwhelming in non-adversarial settings; or they can be applied only to machine learning algorithms that perform poorly for cyber security. We present an original methodology for countering adversarial perturbations targeting intrusion detection systems based on random forests. As a practical application, we integrate the proposed defense method in a cyber detector analyzing network traffic. The experimental results on millions of labelled network flows show that the new detector has a twofold value: it outperforms state-of-the-art detectors that are subject to adversarial attacks; it exhibits robust results both in adversarial and non-adversarial scenarios.
\end{abstract}


\IEEEpeerreviewmaketitle

\section{Introduction}
\label{sec:introduction}

\IEEEPARstart{T}{he} adoption of machine learning to support security operators is an inevitable trend because of the continuous increment of network traffic and sophistication of the attacks~\cite{Buczak:Survey,Blanzieri:Survey,Kettani:Threats}. Machine learning algorithms are employed with success in an increasing number of areas including image processing, speech and text recognition, social media marketing~\cite{Jordan:Machine} and, more recently, in cyber security. Indeed, modern Network Intrusion Detection Systems (NIDS) are being increasingly enriched with machine learning (e.g.,~\cite{Sommer:Outside, Buczak:Survey, Apruzzese:Deep}) and deep learning algorithms (e.g.,~\cite{Apruzzese:Deep, Torres:Analysis, Kim:Long}). Even some commercial products (e.g. Darktrace or Dragonfly Threat Sensor) integrate detectors based on machine learning. 
Despite these positive achievements, recent literature (e.g.,~\cite{Papernot:Limitations, Biggio:Evasion, Papernot:SoK, Su:One}) highlights that existing machine learning techniques are vulnerable to the so called \emph{adversarial attacks}. These malicious actions involve the production of samples designed to thwart the machine learning algorithm by inducing outputs favorable to the attacker. Similar vulnerabilities are critical in the cyber security domain because any undetected attack may compromise an entire organization. The problem of adversarial attacks against machine learning detectors is a relevant open issue.

We propose a novel approach for hardening cyber detectors based on machine learning. We focus on the random forest algorithm due to its proven effectiveness for intrusion detection~\cite{Choudhury:Comparative, Fajana:Torbot, Stevanovic:Efficient, Resende:Survey, Ashfaq:Fuzziness, Apruzzese:Addressing}; however, recent studies also highlight its vulnerability to adversarial perturbations~\cite{Hu:Generating, Apruzzese:Evading, Abaid:Quantifying}. Our solution is based on the observation that existing machine learning cyber detectors rely on excessively rigid classification criteria: they are typically trained through \emph{class labels} that separate samples in disjointed categories where each sample may be either malicious or benign. A similar approach cannot work in the cyber domain where each sample may present more vague attributes. For this reason, we leverage the idea of introducing some degree of flexibility in the training data set by using \emph{probability labels}. The intuition is that a model that uses probability labels instead of hard class labels can be more resilient to adversarial perturbations, and can achieve comparable or even superior results even in the absence of attacks. 
Our methodology has several applications in all fuzzy scenarios characterizing cyber security that involve classifiers based on random forests. As a first test case, in this paper we adopt it for devising botnet detectors based on network flows analyzers.

We validate our approach through a large set of experiments, performed on a set of publicly available and labelled traffic traces containing over $20$ million network flows with benign and malicious samples of different malware families. These data sets capture the network behavior of medium-large enterprises and represent an appropriate setting for a realistic evaluation.
The experimental results demonstrate that the proposed solution devises a detector with comparable or superior performance than state-of-the-art methods in scenarios that are not subject to adversarial attacks. Moreover, it significantly improves the robustness of random forest models against adversarial attacks. Achieving both results is a fundamental success for real contexts where we cannot anticipate whether a machine learning detector will be subject or not to adversarial attacks. Our promising results have room for further improvements, but we are confident that this paper represents a first important step towards more robust cyber defensive platforms based on machine learning against adversarial attacks.

The remainder of this paper is structured as follows. Section~\ref{sec:related} introduces adversarial attacks and compares our paper against related work. Section~\ref{sec:solution} describes the proposed method. Section~\ref{sec:scenario} illustrates the scenario and the threat model considered in this paper. Section~\ref{sec:methodology} presents the methodology and testbeds used for performance evaluation. Section~\ref{sec:experiments} discusses the experimental results. Section~\ref{sec:conclusions} concludes the paper with some final remarks and possible extensions of this work.

\section{Related work}
\label{sec:related}
The complexity of network attacks and the augment of daily traffic requires security operators to rely on some machine learning support~\cite{Buczak:Survey,Blanzieri:Survey}. These methods may detect anomalies and may even reveal attack variants that are not recognizable through signature-based approaches~\cite{Sommer:Outside,Alazab:Zero}. However, the success of novel defensive methods also induce the formulation of new offensive strategies. Today, the so called \textit{adversarial attacks} represent a major limitation to the adoption of a fully autonomous cyber defence platform. We describe the main characteristics of adversarial attacks, and then compare our proposal with the state-of-the-art.

Adversarial attacks are based on the generation of specific samples that induce a machine learning model to produce an output that is favorable to the attacker. This result is caused by the intrinsic sensitivity of machine learning models to their internal configuration settings~\cite{Buczak:Survey, Mannino:Classification, Witten:Data}. Early examples of adversarial attacks against spam filtering are proposed in~\cite{Dalvi:Adversarial,Lowd:Adversarial,Zhou:Combating}. These papers show that linear classifiers could be tricked by few carefully crafted changes in the text of spam emails without affecting the readability of the spam message. Another interesting example of adversarial attack against neural networks classifiers for image processing is presented in~\cite{Szegedy:Intriguing}, where imperceptible perturbations to images used in the training phase can modify arbitrarily the model's output. 
Adversarial attacks can be classified through the taxonomy inspired by~\cite{Huang:Adversarial} that considers the following two properties.

\noindent \textbf{Influence} determines whether an attack is performed at training-time or test-time. 
\begin{itemize}
    \item \textit{Training-time}: these attacks, also known as poisoning attacks, manipulate the training dataset through the insertion or removal of specific samples, therefore altering the decisions of the trained model.
    \item \textit{Test-time}: these attacks subvert the behavior of the detector through the injection of specific samples during its operational phase.
\end{itemize}

\noindent \textbf{Violation} denotes the type of security violation, which can affect the availability or integrity of the system.
\begin{itemize}
    \item \textit{Integrity}: often referred to as evasion attacks, the goal is increasing the false negative rate of the model by introducing malicious samples that are classified as benign.
    \item \textit{Availability}: these attacks tend to cause overwhelming spikes of false alarms, inducing temporary shut-downs and/or recalibrations of the detector.
\end{itemize}

There is extensive literature on adversarial perturbations against image processing (e.g.,~\cite{Papernot:Limitations,Papernot:SoK,Su:One,Jeun:Practical}), while few papers consider adversarial attacks from a cyber security perspective (e.g.,~\cite{Gardiner:Malware,Huang:Adversarial,Hu:Generating, Abaid:Quantifying, Apruzzese:Addressing}). Several recent results demonstrate that adversarial attacks can represent a dangerous threat to any defensive system based on machine learning. For example, \cite{Biggio:Evasion} and~\cite{Xu:Evading} consider the case of adversarial samples against PDF malware detectors based on Support Vector Machines (SVM), neural networks, and random forests. Other papers~\cite{Demontis:Security,Zhang:Adversarial,Abaid:Quantifying} highlight the problem of adversarial evasion for Android malware and spam detectors. Furthermore, the capability of a Generative Adversarial Network to thwart a Domain Generation Algorithm detector based on random forests is evaluated in \cite{Anderson:DeepDGA}. More recently, \cite{Apruzzese:Evading} shows the fragility of a flow-based botnet detector relying on random forest against small adversarial perturbations. Although the threats posed by adversarial inputs are clear, the few existing solutions are not immediately applicable to real contexts. For example,~\cite{Anderson:DeepDGA} proposes to harden the classifier through multiple re-training steps based on adversarial samples. This is an interesting theoretic solution with practical limitations because it requires the creation and continuous management of datasets with realistic adversarial samples. Moreover, \cite{Gardiner:Malware} suggests to improve the robustness against evasion attacks by not considering the features that can be manipulated by an attacker. The problem of this approach is that it reduces accuracy in normal scenarios as shown in~\cite{Demontis:Yes,Apruzzese:Addressing}. On the other hand, our proposal is immediately applicable to real contexts as demonstrated by multiple experimental settings.

Defensive distillation may work in mitigating adversarial perturbations against image classification~\cite{Papernot:Distillation}, but this technique is built and evaluated only on neural network algorithms~\cite{Ross:Improving}. Although cyber detectors based on this algorithm exist and can be hardened through the original distillation proposal~\cite{Grosse:Adversarial}, in cybersecurity scenarios detectors based on random forests outperform those relying on neural networks and other supervised methods~\cite{Stevanovic:Efficient, Resende:Survey, Ashfaq:Fuzziness, Apruzzese:Addressing, Wang:Malware, Choudhury:Comparative}. More recently, \cite{Abraham:Comparison} evaluates different classifiers for the specific problem of botnet detection and confirms that random forest yields the best results. Finally, \cite{Stevanovic:Detecting} proposes a NIDS that inspects network flows through a random forest classifier to identify botnets and obtains outstanding results with detection rates close to $0.99$.
For this reason, we devise an original formulation of the distillation technique that is specifically aimed at hardening random forest detectors, thus allowing to devise robust defensive schemes for cyber detection based on machine learning. 
Although a recent work~\cite{Carlini:Towards} shows that it is possible to evade the defensive distillation, we observe that the considered threat model is unrealistic because it assumes an attacker with complete control of the detector: with similar privileges, attackers can (and most likely will) adopt measures much more invasive and disruptive than those based on adversarial perturbations.
Other works on defenses against adversarial samples~\cite{Biggio:Pattern,Laskov:Practical} consider just SVM classifiers applied to malware analysis, which is out of the scope of this paper. We are not aware of other defensive mechanisms against evasion adversarial attacks that are applicable to random forest algorithms for network intrusion detection. Hence, we can conclude that the topic considered in this paper is a promising research theme, which we address through a novel approach that hardens random forest-based detectors through an original defensive distillation method.

\section{Proposed method}
\label{sec:solution}
We propose a novel method that hardens machine learning detectors based on random forest against adversarial attacks. The idea comes from the observation that the excessively rigid classification criteria learned by machine learning algorithms in the training phase are vulnerable to subtle adversarial perturbations.
Indeed, existing detectors are trained through class labels that separate samples in disjointed categories where each sample may be either malicious or benign but not both. On the other hand, the cyber domain is more fuzzy, and a sample may present characteristics belonging to different categories. Any rigid classification produced by \textit{hard class labels} may represent an exploitable weakness of cyber detectors in adversarial settings. For this reason, we aim to introduce some degree of flexibility and uncertainty in the training process by using \textit{probability labels} that allow the algorithm to capture additional information between classes such as similarity. The intuition is that a model that uses probability labels instead of hard class labels can be more resilient to adversarial samples, and can achieve comparable or superior results even in the absence of attacks. The main difficulty of a similar approach is that probability labels are not readily available in the cyber domain; hence we devise an original solution built upon the two following phases:
\begin{enumerate}
    \item generation of probability labels from hard class labels;
    \item deployment of a supervised model trained with the generated probability labels to perform the cyber detection.
\end{enumerate}

\begin{figure*}[!htbp]
    \centering
    \includegraphics[width=1.5\columnwidth]{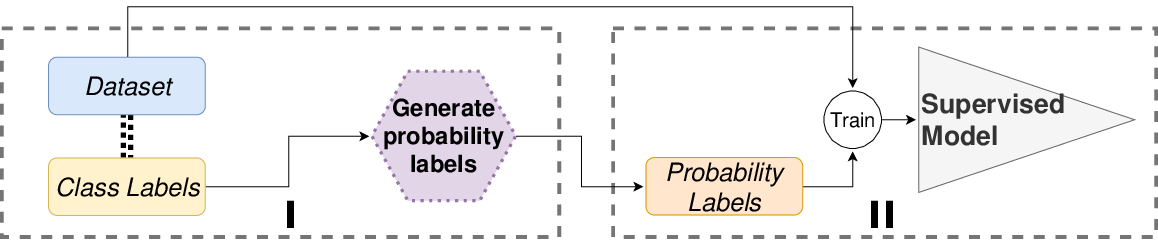}
    \caption{The two phases of the cyber detector.}
    \label{fig:overview}
\end{figure*}

\begin{figure*}[!htbp]
    \centering
    \includegraphics[width=1.8\columnwidth]{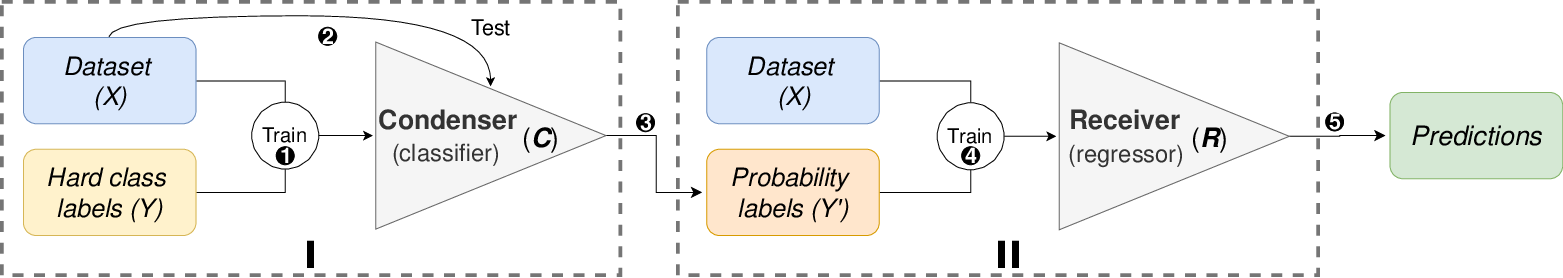}
    \caption{Workflow of the algorithm where distillation is applied to the random forest algorithm.}
    \label{fig:schema}
\end{figure*}

Fig.~\ref{fig:overview} shows that this approach considers as its input a \emph{dataset} and its \emph{class labels}. Then, it computes the corresponding \emph{probability labels} (represented in the leftmost box), and uses them to train a \emph{supervised model} that will be integrated in the detector. We apply this method to the random forest machine learning algorithm by leveraging the foundations~\cite{Pan:Transfer} of the defensive distillation for neural networks~\cite{Papernot:Distillation}. By using the information encoded in the probability labels in the form of probability vectors, generated after training an initial model, it is possible to develop a second ``distilled'' model that is more robust against adversarial attacks. The entire workflow applied to the random forest algorithm is illustrated in Fig.~\ref{fig:schema} where each step is denoted by a circled number that is explained in the following subsections.
Unlike the original defensive distillation technique, the generation of probability labels and their use for detection is performed through random forest-based models instead of neural networks.

\subsection{Generation of the probability labels}
\label{sec:labelGeneration}

The initial phase is performed through a random forest classifier, the \textbf{Condenser}, denoted by~$\mathcal{C}$. We first train this classifier (step \ballnumber{1} in Fig.~\ref{fig:schema}). Then, we leverage the intrinsic property of the random forest algorithm of being an ensemble method, that is, a composition of several decision trees (or estimators), where the final output is generated after evaluating the response of each individual tree. This characteristic allows us to produce the desired probability vectors by considering the percentage of estimators that predicted a specific result (step \ballnumber{2} in Fig.~\ref{fig:schema}). Formally, let $X$ be a dataset, $|X| \in \mathbb{N}$ the number of samples that constitute $X$, and $x_i \in X (0 \leq i \leq |X|)$ a sample within this dataset; let $Y$ be the set of hard class labels (in the form of indicator vectors) associated to dataset $X$, and $y_i \in Y$ the label associated to $x_i$. If $\mathcal{C}$ is a random forest classifier, then $|\mathcal{C}| \in \mathbb{N}$ is the number of estimators that compose $\mathcal{C}$, and $t_j \in \mathcal{C} (0 \leq j \leq |\mathcal{C}|) $ is a tree of classifier $\mathcal{C}$.
After training $\mathcal{C}$ by means of $X$ (as training dataset) and of $Y$ (as labels), the set of probability labels $Y'$ that can be obtained from $X$ through $\mathcal{C}$ is:
\begin{align}
    Y' =  \bigg\{ y'_i \mid y'_i = \frac{\sum_{j=1}^{|\mathcal{C}|}t_{j}^{i}}{|\mathcal{C}|} \bigg\},
\end{align}
where $y'_i$ is the probability vector corresponding to sample $x_i$, and $t_{j}^{i}$ denotes the output of tree $t_j$ for sample $x_i$, which is an indicator vector.
As an example, let us consider a random forest classifier consisting of $100$ estimators that are trained to solve a binary classification problem (either $0$ or $1$). Now, let us assume that, for a given sample, $31$ estimators predict $0$ and produce the indicator vector ($1,0$), while the remaining $69$ predict $1$ and produce the indicator vector ($0,1$). In this case, although the final output of the classifier is the indicator vector ($0,1$), we generate the binary probability vector ($0.31$, $0.69$) which encodes the output produced by each individual tree. On the other hand, if $69$ estimators predict $0$ and $31$ estimators predict $1$, we would obtain the probability vector ($0.69$, $0.31$).

It should be noted that the objective of the Condenser is to generate accurate probability labels but it does not perform detection. As the focus is on the prediction of every individual estimator, and not on the classification results of the whole random forest classifier, the concept of ``misclassification" does not strictly apply to this phase. For example, let us consider a binary classification scenario where we train the Condenser and then test it to generate the probability labels: it may be possible that, for a sample associated to the label $1$,  $69\%$ of the estimators of the Condenser predict a $0$. This event cannot be considered a misclassification because the output of the Condenser is a probability (e.g., the probability vector ($0.69$, $0.31$)). However, such occurrences may have a detrimental effect in the next phase. To minimize similar risks, we utilize the entire available dataset to both train and test $\mathcal{C}$: this approach would yield the best results as it ensures that each sample is associated to a probability label with the highest degree of confidence.

\subsection{Model deployment}
\label{sec:deployment}

In the second phase, the probability vectors generated by the Condenser (step \ballnumber{3} in Fig.~\ref{fig:schema}) are used as training labels for a random forest \textit{regressor} that uses those probabilities as its training input (step \ballnumber{4} in Fig.~\ref{fig:schema}). We define this model as the \textbf{Receiver} denoted by $\mathcal{R}$. Since this model performs the actual detection tasks (step \ballnumber{5} in Fig.~\ref{fig:schema}), we evaluate it against the adversarial inputs. Hence, it is important that this model is trained by following the best practices (as in~\cite{Witten:Data}) to avoid the risk of overfitting. For example, the training and validation sets should be chosen through appropriate splits of the available dataset.

We remark that the Receiver can be seen as a complex multi-output regressor with the challenging task of multi-target regression~\cite{Borchani:Survey}. However, for the specific scenarios related to cyber detection, it is possible to devise a simpler regressor because the main goal is to analyze network traffic and to identify illegitimate activities. Hence, we can model the case as a binary classification instead of a multi-class problem, in which the algorithm is required to determine only whether a given sample of traffic is malicious or not.
To this purpose, for each data sample, the Condenser needs to generate a single probability value (denoting the likelihood of being a malicious sample) instead of a multi-dimensional probability vector. By considering the binary classification example described in Section~\ref{sec:labelGeneration}, the $31$ estimators of the Condenser that predicted a $0$ would give the value $0$, while the remaining $69$ estimators would produce the value $1$. Thus, the corresponding probability value for the analyzed sample is $0.69$.
These probability values are then used as the labels for the Receiver, whose output is another probability value that can be converted into a discrete number through a rounding operation:

\begin{align}
    P(x_i) = \nint{\mathcal{R}_{x_i}},
\end{align}
where $\mathcal{R}_{x_i}$ is the output of the Receiver $\mathcal{R}$ for the sample $x_i$, and $P(x_i) \in [0,1]$ denotes the final prediction of the distilled model.

\section{Application scenario for the detector}
\label{sec:scenario}

\begin{figure}[!htbp]
    \centering
    \includegraphics[width=\columnwidth]{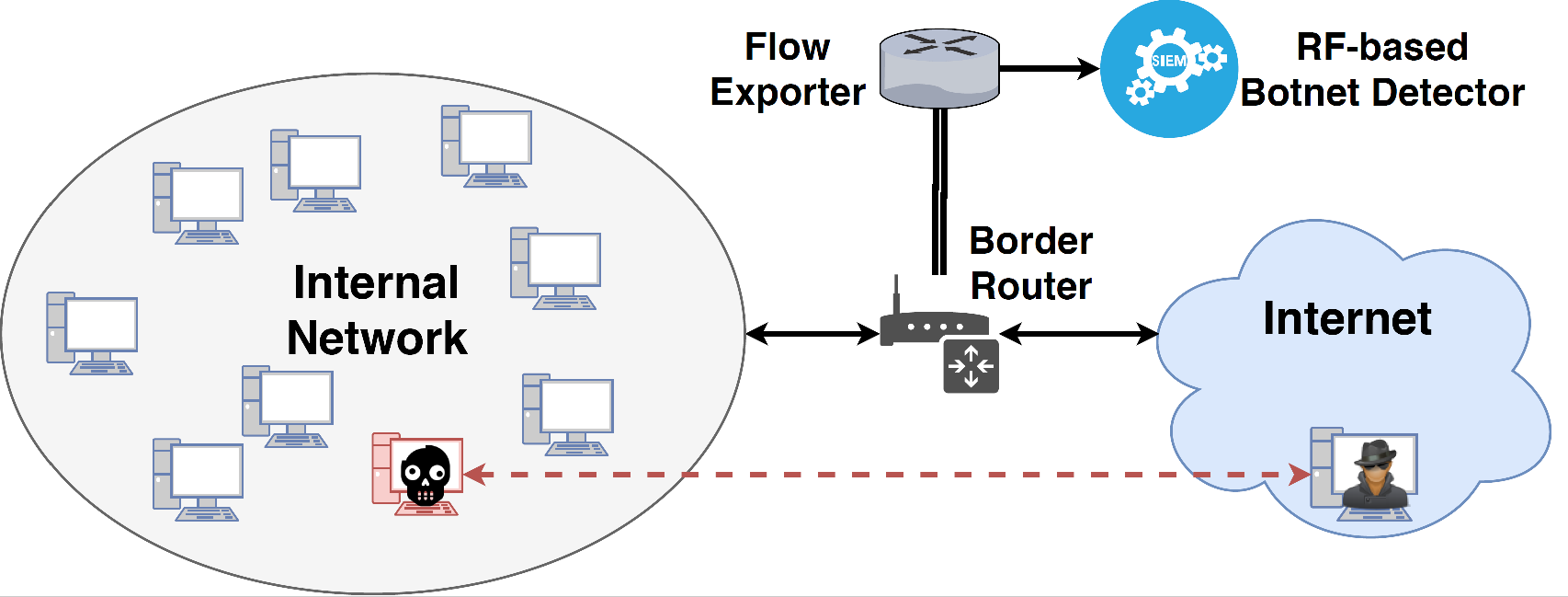}
    \caption{Example of network considered in our use-case.}
    \label{fig:scenario}
\end{figure}

A realistic scenario where the proposed detector can be applied successfully is represented in Fig.~\ref{fig:scenario}, which shows a large enterprise network with many internal hosts and a border router connected to a network flow exporter. The generated flows are inspected by a network intrusion detection system based on machine learning that aims to identify malicious activities (e.g., botnet) by leveraging the random forest algorithm.
We assume that an attacker has already established a foothold in the internal network by compromising one or more machines and deploying botnet malware that communicate with a Command and Control (CnC) infrastructure. The attacker model can be described accordingly to the four characteristics described in~\cite{Biggio:Evasion}: goal, knowledge, capabilities, strategy.

The main goal of the attacker is to evade detection so that he can maintain access to the internal network, compromise more machines and gather information about adopted defenses~\cite{Marchetti:APT}. He knows that network communications are monitored by a NIDS based on machine learning. We assume that the attacker can issue commands to the bot through the CnC infrastructure, possibly modifying the underlying network behavior, but he cannot interact with the detector. Although the attacker does not know the specific machine learning algorithm (alongside its parameters and features) used by the NIDS, he can easily guess that the detector is trained over a dataset containing malicious flows generated by the same or a similar malware variant deployed on the infected machines. Hence, he has to devise some countermeasure to evade the botnet detector.

The strategy to avoid detection is through a \textit{targeted exploratory integrity attack}~\cite{Biggio:Evasion} that is performed by inserting tiny modifications in the communications between the bot and its CnC server.
These alterations may include slight increases of flow duration, exchanged bytes and exchanged packets. Similar changes can be applied without interfering with the application logic of the bot that can continue to operate as initially designed by the attacker. In such a way, the detector is induced to misclassify the network flows generated by bot communications despite being trained with malicious samples belonging to the botnet variant employed by the attacker.

\section{Evaluation methodology}
\label{sec:methodology}

The evaluation and comparison of machine-based detectors subject to adversarial attacks is a complex procedure. In this section, we describe the methodology of our evaluation by presenting the experimental testbed, the details of the considered random forest models, and the procedure to generate the adversarial samples. 

\subsection{Experimental testbed}
\label{sec:testbed}
The experimental evaluation considered in our paper is performed on a public collection of multiple datasets, known as ``CTU-13''~\cite{Garcia:CTU}. The CTU-13 includes network data captured at the Czech Technical University in Prague, and contains labelled network traffic generated by various botnet variants and mixed with normal and background traffic. These flows are captured in a network environment with hundreds of hosts, while the malicious traffic is generated by infecting machines with malware related to several botnet families~\cite{Garcia:CTU}. Overall, the CTU-13 contains $13$ distinct datasets of different botnet activity; each dataset refers to one botnet variant of the $6$ considered families: {\fontfamily{cmtt}\selectfont Neris}, {\fontfamily{cmtt}\selectfont Rbot}, {\fontfamily{cmtt}\selectfont Virut}, {\fontfamily{cmtt}\selectfont Menti}, {\fontfamily{cmtt}\selectfont Murlo}, {\fontfamily{cmtt}\selectfont NSIS.ay}. We report the meaningful metrics of each dataset in the CTU-13 collection in Table~\ref{table:ctu13-metrics}, which also includes the botnet-specific piece of malware and the number of infected machines. This Table highlights the massive amount of included data, which can easily represent the network behavior of a medium-to-large real organization. Nevertheless, we remark that in our evaluation, we prefer not to consider the {\fontfamily{cmtt}\selectfont Sogou} botnet because of the limited amount of its malicious samples.

\begin{table*}[!ht]
\centering
\caption{Meaningful metrics of the CTU-13 collection. Source:~\cite{Garcia:CTU}.}
\begin{tabular}{|c|c|c|c|c|c|c|c|c|}
	\hline
	\textbf{Dataset} & \textbf{Duration (hrs)} & \textbf{Size (GB)} & \textbf{Packets} & \textbf{Netflows} & \textbf{Malicious Flows} & \textbf{Benign Flows} & \textbf{Botnet} & \textbf{\# Bots}  \\
	\hline
	1 & $6.15$ & $52$ & $71\,971\,482$ & $2\,824\,637$ & $40\,959$ & $2\,783\,677$ & {\fontfamily{cmtt}\selectfont Neris} & 1 \\
	\hline
	2 & $4.21$ & $60$ & $71\,851\,300$ & $1\,808\,122$ & $20\,941$ & $1\,787\,181$ & {\fontfamily{cmtt}\selectfont Neris} & 1 \\
	\hline
    3 & $66.85$ & $121$ & $167\,730\,395$ & $4\,710\,638$ & $26\,822$ & $4\,683\,816$ & {\fontfamily{cmtt}\selectfont Rbot} & 1 \\
    \hline
    4 & $4.21$ & $53$ & $62\,089\,135$ & $1\,121\,076$ & $1\,808$ & $1\,119\,268$ & {\fontfamily{cmtt}\selectfont Rbot} & 1 \\
	\hline
    5 & $11.63$ & $38$ & $4\,481\,167$ & $129\,832$ & $901$ & $128\,931$ & {\fontfamily{cmtt}\selectfont Virut} & 1 \\
	\hline
    6 & $2.18$ & $30$ & $38\,764\,357$ & $558\,919$ & $4\,630$ & $554\,289$ & {\fontfamily{cmtt}\selectfont Menti} & 1 \\
	\hline
    7 & $0.38$ & $6$ & $7\,467\,139$ & $114\,077$ & $63$ & $114\,014$ & {\fontfamily{cmtt}\selectfont Sogou} & 1 \\
	\hline
    8 & $19.5$ & $123$ & $155\,207\,799$ & $2\,954\,230$ & $6\,126$ & $2\,948\,104$ & {\fontfamily{cmtt}\selectfont Murlo} & 1 \\
	\hline
    9 & $5.18$ & $94$ & $115\,415\,321$ & $2\,753\,884$ & $184\,979$ & $2\,568\,905$ & {\fontfamily{cmtt}\selectfont Neris} & 10 \\
	\hline
    10 & $4.75$ & $73$ & $90\,389\,782$ & $1\,309\,791$ & $106\,352$ & $1\,203\,439$ & {\fontfamily{cmtt}\selectfont Rbot} & 10 \\
	\hline
    11 & $0.26$ & $5$ & $6\,337\,202$ & $107\,251$ & $8\,164$ & $99\,087$ & {\fontfamily{cmtt}\selectfont Rbot} & 3 \\
	\hline
    12 & $1.21$ & $8$ & $13\,212\,268$ & $325\,471$ & $2\,168$ & $323\,303$ & {\fontfamily{cmtt}\selectfont NSIS.ay} & 3 \\
	\hline
    13 & $16.36$ & $34$ & $50\,888\,256$ & $1\,925\,149$ & $39\,993$ & $1\,885\,156$ & {\fontfamily{cmtt}\selectfont Virut} & 1\\
	\hline
\end{tabular}
\label{table:ctu13-metrics}
\end{table*}

To generate each dataset, the authors first capture the network data in specific packet-capture (PCAP) files, and then convert them into \textit{network flows}. A network flow (or \textit{netflow}) is essentially a sequence of records, each one summarizing a connection between two endpoints (that is, IP addresses).
The inspection of network flows allows administrators to easily summarize the information of two endpoints, such as the source and destination of traffic, the class of service, and the size of transmitted data. Network flows are of particular interest for cyber security applications because of the following benefits with respect to full packet captures: lower amount of storage space required; faster analyses; reduced privacy concerns due to the absence of packet-specific payloads~\cite{Pierazzi:Online}.

The authors of the CTU-13 convert the raw network packets into network flows by means of Argus, a network audit system. Argus presents a client-server architecture: the server component processes packets (either PCAP files or live packet data) and generates detailed status reports of all the netflows in the packet stream, which are then provided to the dedicated clients. By inspecting the CTU-13, we can assume that the client used by the authors to extract the netflows from each individual PCAP file is \texttt{ra}. The output of this conversion process is a CSV file. The final step is the labeling of each individual network flow: indeed, the authors provide an additional ``Label'' field, which separates legitimate from illegitimate flows. More specifically, benign flows correspond to the \textit{normal} and \textit{background} labels; whereas the \textit{botnet} and \textit{CnC-channel} labels denote malicious samples.

\subsection{Considered detectors}
\label{sec:detectors}

\begin{figure}[!htbp]
    \centering
    \includegraphics[width=0.7\columnwidth]{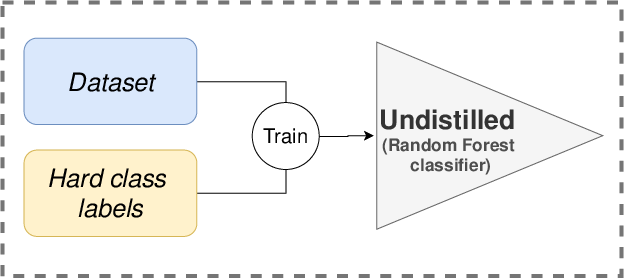}
    \caption{Architecture of the Undistilled detector.}
    \label{fig:undistilled}
\end{figure}

For the evaluation we consider the following detectors based on random forest: 
\begin{itemize}
    \item The \textbf{Undistilled} detector, which presents characteristics similar to the random forest classifier model proposed in~\cite{Stevanovic:Botnet}, is used as the baseline for the experiments; a graphical representation of its architecture is provided in Fig.~\ref{fig:undistilled}.
    \item The \textbf{Distilled} detector represents the main proposal of this paper. It consists of the \textit{Condenser} for generating the probability labels, and of the \textit{Receiver} to perform the detection tasks. This detector is evaluated against the Undistilled detector in adversarial and non-adversarial settings.
\end{itemize}

Each detector has $6$ instances, each one focusing on recognizing a specific malware family of the dataset. The motivation for this design choice comes from the observation that machine learning techniques yield superior results when they pursue a specific goal rather than aiming to an impossible catch-all solution~\cite{Apruzzese:Deep,Stevanovic:Botnet}. 

For each botnet variant, we generate a dedicated training set containing both benign and malicious samples belonging to that family; all instances share the same legitimate-to-illegitimate flow ratio in the training sets. Formally, let $D$ be the set of all the traces of network flows considered in the testbed, and let $D^l \subset D$ and $D^m \subset D$ be the sets of all legitimate and malicious samples in $D$, respectively (so that $D^l \cup D^m = D$, and $D^l \cap D^m = \emptyset$). Now, let $D^b$ be the set of malicious flows corresponding to the $b$ botnet family, so that $\bigcup_{b=1}^{6} D^b = D^m$. We train each detector's instance corresponding to the $b$ botnet family with samples randomly extracted from $D^l$ and $D^b$, in a $20:1$ ratio. (The randomized extraction of samples is done to reduce the impact of selection bias.) The $20:1$ ratio is similar to that in~\cite{Stevanovic:Efficient}, and it is motivated by the fact that in realistic settings the legitimate flows largely outnumber the botnet-generated flows. Other studies use even greater ratios~\cite{Kirubavathi:Botnet}. The instances of the Receiver are trained with $80\%$ of the botnet flows generated by each malware variant, and validated on the remaining $20\%$. These splits are close to those adopted in~\cite{Stevanovic:Efficient, Apruzzese:Evading}. On the other hand, the instances of the Condenser, which generate the probability labels, are trained and tested on the same dataset containing all the malicious flows of the related botnet family. Other details are presented in Section~\ref{sec:solution}. 
These models adopt feature sets that are similar to those adopted in~\cite{Apruzzese:Evading} and~\cite{Stevanovic:Botnet} because they achieve appreciable detection rates. We integrate these features with information about the IANA \textit{port type} for the source and destination hosts, thus obtaining the list summarized in Table~\ref{tab:features}. For completeness, we remark that the code for the experiments is implemented in \textit{Python3} and uses the \textit{scikit-learn} toolkit. Moreover, we report in Table~\ref{tab:parameters} the meaningful parameter settings of each model, which are chosen through extensive grid search operations. The $F$ parameter denotes the number of features in input, and MSE is the Mean Squared Error.

\begin{table}[h]
\centering
\caption{Features of the random forest models. Source:~\cite{Apruzzese:Evading}.}
\resizebox{!}{0.75\columnwidth}{
    \begin{TAB}[2pt]{|c|c|c|}{|c|c|c|c|c|c|c|c|c|c|c|c|c|c|c|c|}
    \centering
    \textbf{\#} & \textbf{Feature name} & \textbf{Feature type}  \\
    1,2 & source/destination IP address type & Boolean \\
    3,4 & source/destination port & Numerical \\
    5 & flow direction & Boolean \\
    6 & connection state & Categorical \\
    7 & duration (seconds) & Numerical \\
    8 & protocol & Categorical \\
    9,10 & source/destination ToS & Numerical \\
    11,12 & outgoing/incoming bytes & Numerical \\
    13 & total transmitted packets & Numerical \\
    14 & total transmitted bytes & Numerical \\
    15,16 & source/destination port type & Categorical \\
    17 & bytes per second & Numerical \\
    18 & bytes per packet & Numerical \\
    19 & packets per second & Numerical\\
    20 & ratio of outgoing/incoming bytes & Numerical \\
    \end{TAB}
}
\label{tab:features}
\end{table}

\begin{table}[h]
    \centering
    \caption{Parameters of the random forest models.}
    \resizebox{!}{65px}{%
        \begin{tabular}{c||c|c|c}
            \multicolumn{1}{c}{} & \multicolumn{1}{c}{\textbf{Parameter name}} & \multicolumn{1}{c}{\textbf{Value}} \\
            \toprule
        
            \parbox[t]{2mm}{\multirow{4}{*}{\rotatebox[origin=c]{90}{\textbf{Undistilled}}}} & Number of estimators & $763$ \\
            & Quality Function & Gini \\
            & Features for best split & $\sqrt{F}$ \\
            & Bootstrap & Yes \\
            \midrule
            
            \parbox[t]{2mm}{\multirow{4}{*}{\rotatebox[origin=c]{90}{\textbf{Condenser}}}} & Number of estimators & $894$ \\
            & Quality Function & Gini \\
            & Features for best split & $\sqrt{F}$ \\
            & Bootstrap & Yes \\
            \midrule
            
            \parbox[t]{2mm}{\multirow{4}{*}{\rotatebox[origin=c]{90}{\textbf{Receiver}}}} & Number of estimators & $1352$ \\
            & Quality Function & MSE\\
            & Features for best split & $F/2$ \\
            & Bootstrap & Yes \\
            \bottomrule
        \end{tabular}
    }
    \label{tab:parameters}
\end{table}

\subsection{Generation of adversarial datasets}
\label{sec:generation}
We produce multiple adversarial datasets by manipulating the botnet netflows $D^b$ through feature modifications. Since the produced adversarial samples are used to evaluate the proposed approach, we consider the portion of botnet netflows from $D^b$ contained in the datasets used for the testing-phase, thus avoiding the submission of samples contained in the training set.

An attacker can evade detection by increasing the flow duration through a small latency; and the number of bytes (or packets) by adding random junk data. All these modifications can be introduced in the network behavior of the bots without altering their underlying logic. To reproduce a similar adversarial attack pattern, we generate adversarial samples by manipulating combinations of up to $4$ features, such as the duration of the flows, the total number of transmitted packets, the number of outgoing(Src) or incoming(Dst) bytes. Table~\ref{tab:combinations} reports the $15$ groups of altered features denoted by $G$. As an example, adversarial samples belonging to group {\fontfamily{qcr}\selectfont 1a} alter only the flow \textit{duration}, while those of group {\fontfamily{qcr}\selectfont 3c} include modifications to the \textit{duration}, \textit{dst\_bytes} and \textit{tot\_packets} features.
The feature manipulation is performed by augmenting each of these groups through $9$ increment steps denoted by $S$; these steps are fixed for all the possible combinations. Hence, for each botnet family, we produce $135$ adversarial collections, thus resulting in a total of $810$ adversarial datasets (given by $15 [\text{groups of altered features}] * 9 [\text{increment steps}] * 6 [\text{botnet families}]$).

\begin{table}[!ht]
\centering
\caption{Groups of altered features. Source:~\cite{Apruzzese:Evading}.}
\resizebox{0.8\columnwidth}{!}{
	\begin{tabular}{c|c}
		\toprule
		\textbf{Group} (g) & \textbf{Altered features}  \\
		\midrule
		{\fontfamily{qcr}\selectfont \bfseries 1a} & Duration (in seconds)\\
    {\fontfamily{qcr}\selectfont \bfseries 1b} & Src\_bytes \\
    {\fontfamily{qcr}\selectfont \bfseries 1c} & Dst\_bytes \\
    {\fontfamily{qcr}\selectfont \bfseries 1d} & Tot\_pkts \\
    {\fontfamily{qcr}\selectfont \bfseries 2a} & Duration, Src\_bytes \\
    {\fontfamily{qcr}\selectfont \bfseries 2b} & Duration, Dst\_bytes \\ 
    {\fontfamily{qcr}\selectfont \bfseries 2c} & Duration, Tot\_pkts \\ 
    {\fontfamily{qcr}\selectfont \bfseries 2d} & Src\_bytes, Tot\_pkts \\
	{\fontfamily{qcr}\selectfont \bfseries 2e} & Src\_bytes, Dst\_bytes \\ 
    {\fontfamily{qcr}\selectfont \bfseries 2f} & Dst\_bytes, Tot\_pkts \\
    {\fontfamily{qcr}\selectfont \bfseries 3a} & Duration, Src\_bytes, Dst\_bytes \\
    {\fontfamily{qcr}\selectfont \bfseries 3b} & Duration, Src\_bytes, Tot\_pkts \\
    {\fontfamily{qcr}\selectfont \bfseries 3c} & Duration, Dst\_bytes, Tot\_pkts \\
    {\fontfamily{qcr}\selectfont \bfseries 3d} & Src\_bytes, Dst\_bytes, Tot\_pkts \\
    {\fontfamily{qcr}\selectfont \bfseries 4a} & Duration, Src\_bytes, Dst\_bytes, Tot\_pkts \\
		\bottomrule
	\end{tabular}
}
\label{tab:combinations}
\end{table}

Table~\ref{tab:increments} reports the relationship between each step and the corresponding feature increments where \textit{Duration} is measured in seconds. As an example, the adversarial datasets obtained through the {\fontfamily{qcr}\selectfont VI} step of the group {\fontfamily{qcr}\selectfont 1b} have the values of their flow outgoing bytes increased by $128$. The adversarial datasets obtained through the {\fontfamily{qcr}\selectfont II} step of the group {\fontfamily{qcr}\selectfont 3c} have the values of their flow duration, incoming bytes and total packets increased by $2$. There is a greater focus on small increments since they are easier to achieve and they are still able to generate samples that evade detection. The rationale behind the choice of the values shown in Table~\ref{tab:increments} is the following: our objective is to generate adversarial malicious samples that are only marginally different from their original counterparts, as shown in~\cite{Su:One}. Although the exact numbers have been selected arbitrarily by adopting the powers of $2$ for convenience, our goal is to represent the effects of small, but sensible variations of these features. Furthermore, introducing these small perturbations is a realistic task for the type of attacker considered in this paper. On the other hand, excessive increases higher than those shown in Table~\ref{tab:increments} may generate anomalous network flows that can be detected by different defensive mechanisms (e.g.,~\cite{Pierazzi:Online}). Moreover, increasing the duration of each flow above 120 seconds may exceed the duration limits of the flow collector~\cite{Pierazzi:Online}. 

\begin{table}[!ht]
\centering
\caption{Increment steps of each feature for generating realistic adversarial samples. Source:~\cite{Apruzzese:Evading}.}
\resizebox{0.8\columnwidth}{!}{
	\begin{tabular}{c|cccc}
		\hline
		\textbf{Step} (s) & \textbf{Duration} & \textbf{Src\_bytes} & \textbf{Dst\_bytes} & \textbf{Tot\_pkts} \\
		\hline
		{\fontfamily{qcr}\selectfont \bfseries I} & $+1$ & $+1$ & $+1$ & $+1$ \\
    \hline
    {\fontfamily{qcr}\selectfont \bfseries II} & $+2$ & $+2$ & $+2$ & $+2$ \\
    \hline
    {\fontfamily{qcr}\selectfont \bfseries III} & $+5$ & $+8$ & $+8$ & $+5$ \\
    \hline
    {\fontfamily{qcr}\selectfont \bfseries IV} & $+10$ & $+16$ & $+16$ & $+10$ \\
    \hline
    {\fontfamily{qcr}\selectfont \bfseries V} & $+15$ & $+64$ & $+64$ & $+15$ \\
    \hline
    {\fontfamily{qcr}\selectfont \bfseries VI} & $+30$ & $+128$ & $+128$ & $+20$ \\
    \hline
    {\fontfamily{qcr}\selectfont \bfseries VII} & $+45$ & $+256$ & $+256$ & $+30$ \\
    \hline
    {\fontfamily{qcr}\selectfont \bfseries VIII} & $+60$ & $+512$ & $+512$ & $+50$ \\
    \hline
    {\fontfamily{qcr}\selectfont \bfseries IX} & $+120$ & $+1024$ & $+1024$ & $+100$ \\
    \hline
	\end{tabular}
}
\label{tab:increments}
\end{table}

The generation of the adversarial datasets is described in Algorithm~\ref{alg:generateAdversarialSamples}, where $\mathcal{A}(\cdot)$ denotes the operator indicating an adversarially manipulated input. We remark the importance of the operation on line~\ref{alg:update}, because it shows that some features are mutually dependent. For example, for consistency reasons, increasing the flow duration requires to update also the bytes per second and the packets per second.

For the performance evaluation we adopt the typical machine learning metrics: \emph{Precision} (\textit{Prec}), \emph{Detection Rate} (\emph{DR}, or \emph{Recall}), \emph{F1-score}, computed as follows:
\vspace*{-2mm}
\begin{tabularx}{1\columnwidth}{@{}XX@{}}
  \begin{equation}  Prec = \frac{TP}{TP+FP},   \end{equation} & \begin{equation}   DR = \frac{TP}{TP+FN}, \end{equation}
\end{tabularx}
\vspace*{-1mm}
\begin{equation}
    F1{\text -}score = 2*\frac{Precision * DR}{Precision+DR},
\end{equation}
\noindent
where $TP$, $FP$, and $FN$ denote true positives, false positives, and false negatives, respectively. In the remainder of this paper, we consider a positive detection as a malicious sample.

\begin{algorithm2e}
    \caption{Algorithm for generating datasets of adversarial samples.}
    \footnotesize
    \label{alg:generateAdversarialSamples}
    \DontPrintSemicolon
    \SetAlgoNoEnd
    
    \KwIn{List of datasets of malicious flows $X^m$ divided in botnet-specific sets $X^b$; list of altered features groups $G$; list of feature increment steps $S$.}
    \KwOut{List of adversarial datasets $\mathcal{A}(X^m)$.}
    
    \hrule
    $\mathcal{A}(X^m)$ $\gets$ emptyList();\\
    
    \ForEach{$group$ $g \in G$}{
        \ForEach{$step$ $s \in S$}{
            \ForEach{$dataset$ $X^b \in X^m$}{
                $\mathcal{A}^{g}_s(X^b)$ $\gets$ $CreateOneDataset(s,g,X^b)$;\\
                Insert $\mathcal{A}^{g}_s(X^b)$ in $\mathcal{A}(X^m)$;
            }
        }
    }
    \Return $\mathcal{A}(X^m)$
    
    {\tt \scriptsize // Function for creating a single adversarial dataset $\mathcal{A}^{g}_s(X^b)$ corresponding to a botnet-specific dataset $X^b$, a specific altered feature group $g$, and a specific increment step $s$.}\\
    
    \SetKwFunction{proc}{proc}
    \SetKwProg{myproc}{Function}{}{}
    \myproc{CreateOneDataset($s, g, X^b$)}{ \label{alg:oneDataset}
        $\mathcal{A}^{g}_s(X^b)$ $\gets$ emptyList();\\
        \ForEach{$sample$ $x^b \in X^b$}{
            $\mathcal{A}^{g}_s(x^b)$ $\gets$ $AlterSample(s,g,x^b)$;\\
            Insert $\mathcal{A}^{g}_s(x^b)$ in $\mathcal{A}^{g}_s(X^b)$;
        }
        \Return $\mathcal{A}^{g}_s(X^b)$
    }
        
    {\tt \scriptsize // Function for creating a single adversarial sample $\mathcal{A}^{g}_s(x^b)$ corresponding to a botnet-specific sample $x^b$, a specific altered feature group $g$, and a specific increment step $s$.}\\
    
    \SetKwFunction{proc}{proc}
    \SetKwProg{myproc}{Function}{}{}
    \myproc{AlterSample($s, g, x^b$)}{ \label{alg:oneSample}
        $\mathcal{A}^{g}_s(x^b)$ $\gets$ $x^b$;\\
        Increment features $g$ of $\mathcal{A}^{g}_s(x^b)$ by $s$;\\
        Update features of $\mathcal{A}^{g}_s(x^b)$ that depend on $g$;\\ \label{alg:update}
        \Return $\mathcal{A}^{g}_s(x^b)$
    }

\end{algorithm2e}

\section{Performance evaluation}
\label{sec:experiments}
We present the results of a large set of experiments with the aim of demonstrating that: (i) the proposed distilled random forest detector achieves comparable or better detection performance than state-of-the-art algorithms in scenarios that are not subject to adversarial inputs; (ii) it significantly improves the robustness of machine learning models against adversarial attacks. Achieving both results is an important outcome for cyber security contexts where we cannot anticipate whether a machine learning detector is subjected or not to adversarial attacks.

We evaluate and compare the performance of distilled and undistilled models in scenarios where samples are not adversarially modified. Then, we assess the effectiveness of the distilled random forest model against adversarial perturbations. Finally, we compare the result of the proposed method against two existing defensive strategies that can be applied to any supervised machine learning algorithm.

\subsection{Evaluation in normal scenarios}
\label{sec:normal}

We initially generate the probability labels for the Distilled detector by training and testing its Condenser model. Then, we train both the Distilled (through the Receiver) and Undistilled detectors on the same training set (but with appropriate labels), and proceed to evaluate them on the same test set. The results are shown in Table~\ref{tab:performance}, where the columns report the chosen evaluation metrics, and the rows denote the botnet-specific instances of the Undistilled and Distilled detectors; the last row summarizes the results of each detector, which are averaged among all instances. From this table, we observe that the Distilled detector achieves the best results as it obtains higher Precision and F1-scores, and superior detection rates. We stress that the performance of the Distilled is similar to that obtained by state-of-the-art random forest-based botnet detectors~\cite{Abraham:Comparison,Stevanovic:Botnet}. Furthermore, we highlight that our proposal also outperforms the initial defensive distillation technique applied to neural networks in non-adversarial settings, because the distilled neural network model presents a reduced accuracy of \texttildelow$1.5\%$ when compared to a not-distilled neural network model~\cite{Papernot:Distillation}; this performance drop also affects distilled neural networks for malware classification scenarios~\cite{Grosse:Adversarial}, which exhibit an increased rate of false alarms. It is important to note that the unusual perfect $Prec$ scores achieved by both models for the {\fontfamily{cmtt}\selectfont Murlo} botnet and by the Undistilled model for the {\fontfamily{cmtt}\selectfont Menti} botnet can be motivated as follows: the large majority of the network flows generated by these botnet variants are significantly different from benign traffic, hence the models are able to recognize their malicious samples without generating false positives; however, some instances are still able to evade detection as indicated by the imperfect Recall value.
These experiments show that, in the absence of adversarial attacks, our version of the distillation technique applied to random forests yields a detector with similar or superior performance than those that do not adopt a distillation technique. These results are crucial because they refer to a large set of scenarios and demonstrate that random forest-based detectors integrated with distillation are effective even in the absence of adversarial inputs. 

\begin{table}
    \centering
    \caption{Baseline vs. Distilled model performance.}
    \resizebox{0.85\columnwidth}{!}{
    \begin{tabular}{|c|c|c|c|c|}
		\hline
		    \textbf{Botnet} & \textbf{Detector} & \textbf{F1-Score} & \textbf{Precision} & \textbf{Recall} \\ 
		\hline\hline
		    \multirow{2}{*}{\fontfamily{cmtt}\selectfont Neris} 
		    & \cellcolor{gray!45}Undistilled & \cellcolor{gray!45}$0.9577$ & \cellcolor{gray!45}$0.9615$ & \cellcolor{gray!45}$0.9540$ \\
		    \cline{2-5}
		    & \cellcolor{gray!20}Distilled & \cellcolor{gray!20}$0.9651$ & \cellcolor{gray!20}$0.9671$ & \cellcolor{gray!20}$0.9632$ \\ 
    		\hline\hline
    		
            \multirow{2}{*}{\fontfamily{cmtt}\selectfont Virut} 
	        & \cellcolor{gray!45}Undistilled & \cellcolor{gray!45}$0.9682$ & \cellcolor{gray!45}$0.9876$ & \cellcolor{gray!45}$0.9496$ \\
	        \cline{2-5}
            & \cellcolor{gray!20}Distilled & \cellcolor{gray!20}$0.9753$ & \cellcolor{gray!20}$0.9876$ & \cellcolor{gray!20}$0.9633$  \\
		    \hline\hline
		    
            \multirow{2}{*}{\fontfamily{cmtt}\selectfont Murlo} 
		    & \cellcolor{gray!45}Undistilled & \cellcolor{gray!45}$0.9932$ & \cellcolor{gray!45}$1$ & \cellcolor{gray!45}$0.9866$  \\
		    \cline{2-5}
            & \cellcolor{gray!20}Distilled & \cellcolor{gray!20}$0.9968$ & \cellcolor{gray!20}$1$ & \cellcolor{gray!20}$0.9937$  \\
		    \hline\hline
            
            \multirow{2}{*}{\fontfamily{cmtt}\selectfont Rbot} 
		    & \cellcolor{gray!45}Undistilled & \cellcolor{gray!45}$0.9994$ & \cellcolor{gray!45}$0.9999$ & \cellcolor{gray!45}$0.9999$ \\
		    \cline{2-5}
            & \cellcolor{gray!20}Distilled & \cellcolor{gray!20}$0.9995$ & \cellcolor{gray!20}$0.9999$ & \cellcolor{gray!20}$0.9990$ \\
		    \hline\hline
            
            \multirow{2}{*}{\fontfamily{cmtt}\selectfont Menti} 
		    & \cellcolor{gray!45}Undistilled & \cellcolor{gray!45}$0.9984$ & \cellcolor{gray!45}$1$ & \cellcolor{gray!45}$0.9969$ \\
		    \cline{2-5}
            & \cellcolor{gray!20}Distilled & \cellcolor{gray!20}$0.9979$ & \cellcolor{gray!20}$0.9997$ & \cellcolor{gray!20}$0.9969$  \\
		    \hline\hline
            
            \multirow{2}{*}{\fontfamily{cmtt}\selectfont NSIS.ay} 
		    & \cellcolor{gray!45}Undistilled & \cellcolor{gray!45}$0.9213$ & \cellcolor{gray!45}$0.9925$ & \cellcolor{gray!45}$0.8596$ \\
		    \cline{2-5}
            & \cellcolor{gray!20}Distilled & \cellcolor{gray!20}$0.9273$ & \cellcolor{gray!20}$0.9784$ & \cellcolor{gray!20}$0.8812$ \\
		    \hline\hline
            
            \multirow{2}{*}{Average} 
            & \cellcolor{gray!45}Undistilled & \cellcolor{gray!45}$0.9729$ & \cellcolor{gray!45}$0.9774$ & \cellcolor{gray!45}$0.9684$ \\
            \cline{2-5}
            & \cellcolor{gray!20}Distilled & \cellcolor{gray!20}$0.9777$ & \cellcolor{gray!20}$0.9804$ & \cellcolor{gray!20}$0.9751$ \\
		    \hline
		    
    \end{tabular}
    }
    \label{tab:performance}
\end{table}

Since supervised machine learning methods for cyber defense need periodic re-trainings~\cite{Apruzzese:Deep}, it is important to evaluate the computational cost of the proposed solution. Thus, we measure and report the training times of the considered detectors in Table~\ref{tab:times}, which compares the time (in seconds) required for training the baseline Undistilled detector (composed of a single random forest classifier) with those required by our method; as the proposed Distilled detector includes both the Condenser and the Receiver, we report the combined training time of these components. Computations are performed on a machine with the following hardware: CPU Intel Core~i7-7700HQ, RAM 32GB, and SSD 512GB. We observe that training the Distilled detector requires more effort, because it is composed of two models and, in addition, training a random forest regressor (that is, the Receiver) is more demanding than training a classifier. However, we stress that these operations needs to be executed only periodically. Moreover, by performing the training computations on machines with dedicated hardware it is possible to decrease the absolute training time difference to negligible amounts.

\begin{table}[!ht]
\centering
\caption{Training time of each instance of the detectors.}
\resizebox{0.6\columnwidth}{!}{
	\begin{tabular}{|c|c|c|}
		\hline
		\textbf{Botnet} & \textbf{Detector} & \textbf{Time} (s)  \\
		\hline\hline
		
		\multirow{2}{*}{\fontfamily{cmtt}\selectfont Neris}
		    & \cellcolor{gray!45}Undistilled & \cellcolor{gray!45}$75.8$\\ 
		    \cline{2-3} 
		    & \cellcolor{gray!20}Distilled & \cellcolor{gray!20}$212.5$  \\
		\hline\hline
		
	\multirow{2}{*}{\fontfamily{cmtt}\selectfont Virut}
		    & \cellcolor{gray!45}Undistilled & \cellcolor{gray!45}$16.7$\\ 
		    \cline{2-3} 
		    & \cellcolor{gray!20}Distilled & \cellcolor{gray!20}$42.7$ \\
		\hline\hline
		
		\multirow{2}{*}{\fontfamily{cmtt}\selectfont Murlo}
		    & \cellcolor{gray!45}Undistilled & \cellcolor{gray!45}$19.8$ \\ 
		    \cline{2-3} 
		    & \cellcolor{gray!20}Distilled & \cellcolor{gray!20}$53.9$  \\
		\hline\hline
		
		\multirow{2}{*}{\fontfamily{cmtt}\selectfont Rbot}
		    & \cellcolor{gray!45}Undistilled & \cellcolor{gray!45}$77.1$ \\
		    \cline{2-3} 
		    & \cellcolor{gray!20}Distilled & \cellcolor{gray!20}$210.4$  \\
		\hline\hline
		
		\multirow{2}{*}{\fontfamily{cmtt}\selectfont Menti}
		    & \cellcolor{gray!45}Undistilled & \cellcolor{gray!45}$2.8$ \\ 
		    \cline{2-3} 
		    & \cellcolor{gray!20}Distilled & \cellcolor{gray!20}$8.5$  \\
		\hline\hline
		
		\multirow{2}{*}{\fontfamily{cmtt}\selectfont NSIS.ay}
		    & \cellcolor{gray!45}Undistilled & \cellcolor{gray!45}$1.6$ \\ 
		    \cline{2-3} 
		    & \cellcolor{gray!20}Distilled & \cellcolor{gray!20}$5.7$\\
		\hline\hline
		
		\multirow{2}{*}{Average}
		    & \cellcolor{gray!45}Undistilled & \cellcolor{gray!45}$32.3$ \\ 
		    \cline{2-3} 
		    & \cellcolor{gray!20}Distilled & \cellcolor{gray!20}$87.0$ \\
		\hline
	\end{tabular}
}
\label{tab:times}
\end{table}

\subsection{Evaluation in adversarial settings}
It must be determined whether and to which extent the proposed method is able to address issues related to adversarial attacks. To this purpose, we test the Distilled and the Undistilled detectors against the generated adversarial datasets, and compare their performance. The detection rate is the metric of interest for these analyses.
We anticipate that this evaluation highlights a twofold improvement of our proposal: a significant increase in the detection rate; a more stable behavior against different adversarial samples of the same botnet family.

\begin{figure}[!htbp]
    \centering
    \includegraphics[width=0.9\columnwidth]{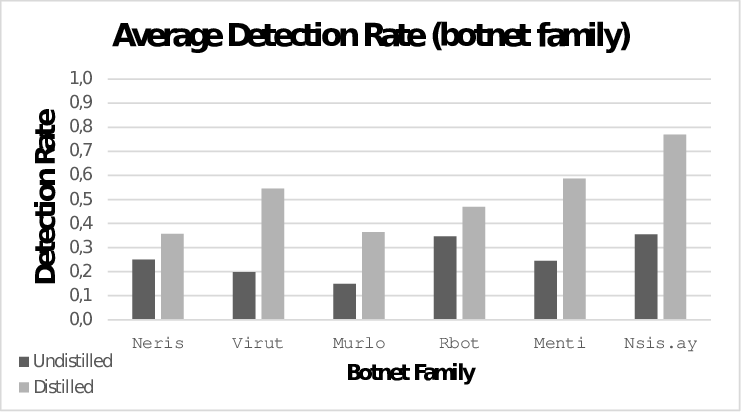}
    \caption{Comparison of the average detection rates on each malware family.}
    \label{fig:avg_botnet}
\end{figure}

Among the considered $810$ adversarial datasets, the Distilled detector clearly outperforms the baseline Undistilled in $759$ cases; for the remaining $51$ datasets, the results of the two detectors are close. 
A comprehensive overview of the effectiveness of the two detectors is presented in Fig.~\ref{fig:avg_botnet}, where the black and gray histograms report the detection rates of the Undistilled and Distilled detectors, respectively. Each histogram denotes the average performance of the models applied to each botnet family. There is no doubt that the Distilled is significantly superior to the Undistilled detector, with improvements ranging from $50\%$ to $250\%$. 

\begin{figure}[!htbp]
    \centering
    \includegraphics[width=0.9\columnwidth]{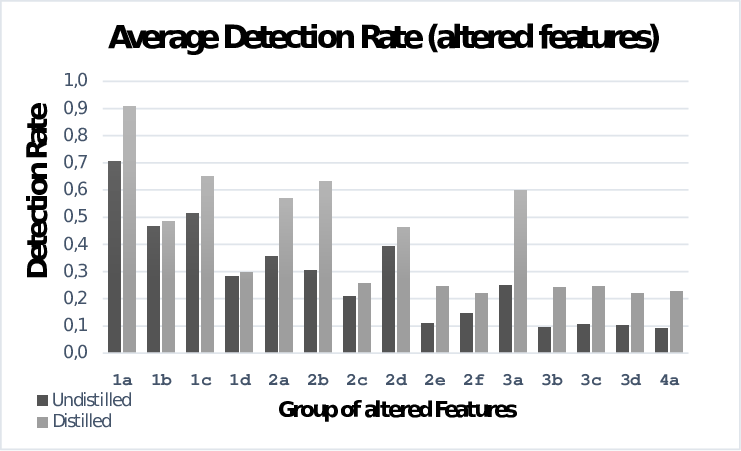}
    \caption{Comparison of the average detection rates for each group of altered features.}
    \label{fig:avg_features}
\end{figure}

We provide a more detailed comparison of the two detectors by considering the impact on detection rates of different altered features. The results are reported in Fig.~\ref{fig:avg_features}, where the x-axis denotes the group of altered features, and every histogram is generated by averaging the detection rates achieved by each instance of the detectors for all increment steps. From this figure, we can observe that the Distilled achieves superior detection rates for all the groups. The improvements for the groups {\fontfamily{qcr}\selectfont 2a}, {\fontfamily{qcr}\selectfont 2b} and {\fontfamily{qcr}\selectfont 3a} are the most significant, as they allow the Distilled to retain a detection rate that is much higher than that of the Undistilled model. Moreover, the results for group {\fontfamily{qcr}\selectfont 1a} show that the Distilled detector is almost unaffected by alterations of the flow duration. On the other hand, adversarial alterations involving multiple features have a high impact on the performance of both detectors, as these modifications cause the malicious test samples to be considerably different than those used to train each model. Nevertheless, it is appreciable that, even in these tough circumstances, the Distilled is able to correctly identify more than twice the amount of malicious flows with respect to the Undistilled detector.

\begin{figure}[!htbp]
    \centering
    \includegraphics[width=0.9\columnwidth]{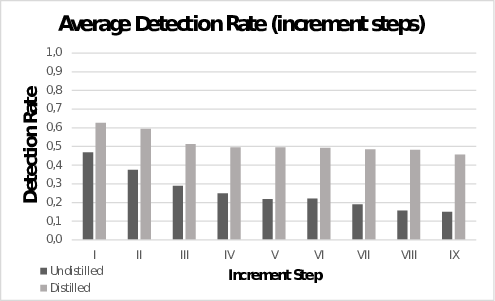}
    \caption{Comparison of the average detection rates for each increment step.}
    \label{fig:avg_increment}
\end{figure}

We also evaluate the detection rates of the two detectors for variable increment steps. The results are presented in Fig.~\ref{fig:avg_increment}, where the x-axis represents the increment steps and the histograms are generated by averaging the performance over all groups of altered features. We note that not only the Distilled outperforms the Undistilled model, but that it is much more resilient against samples that greatly differ from their original malicious version. Indeed, the detection rates for the {\fontfamily{qcr}\selectfont VIII} and {\fontfamily{qcr}\selectfont IX} steps are close to $50\%$; whereas the $15\%$ detection rate of the Undistilled model is unacceptably low. This figure also shows that the Distilled presents a more stable behavior against adversarial samples that are obtained through different increment steps: its detection rates are between $46\%$ and $61\%$, against the much broader $11\%$ to $45\%$ range of the Undistilled model. From Fig.~\ref{fig:avg_increment} and Fig.~\ref{fig:avg_features}, we observe that greater perturbations correspond to the lowest detection rates; however, we remark that such modifications may generate alerts from other defensive mechanisms (as explained in Section~\ref{sec:methodology}). Furthermore, we highlight that adversarial attacks are more effective and more difficult to detect when they are carried out through adversarial samples that are as close as possible to original samples.

\begin{figure*}[!htbp]
    \centering
    
        \begin{subfigure}{\columnwidth}
            \centering
                \includegraphics[width=0.75\linewidth]{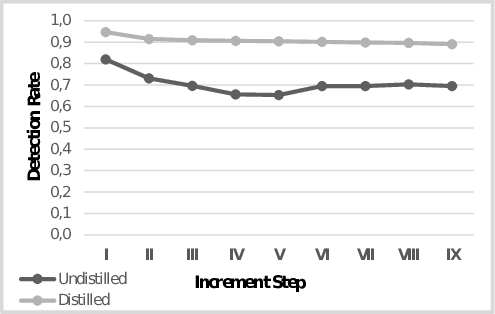}
            \caption{Group {\fontfamily{qcr}\selectfont 1a}: \textit{Duration}.}
            \label{fig:lines1}
        \end{subfigure}%
        \begin{subfigure}{\columnwidth}
            \centering
                \includegraphics[width=0.75\linewidth]{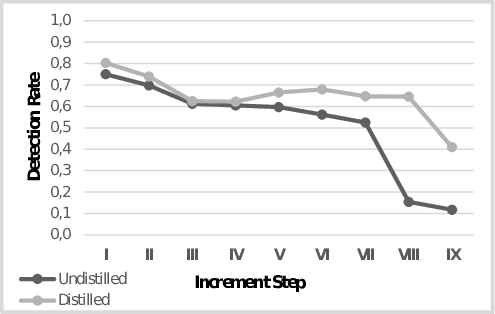}
            \caption{Group {\fontfamily{qcr}\selectfont 1c}: \textit{Dst\_Bytes}.}
            \label{fig:lines2}
        \end{subfigure}
    
        \par\bigskip
        \begin{subfigure}{\columnwidth}
            \centering
            \includegraphics[width=0.75\linewidth]{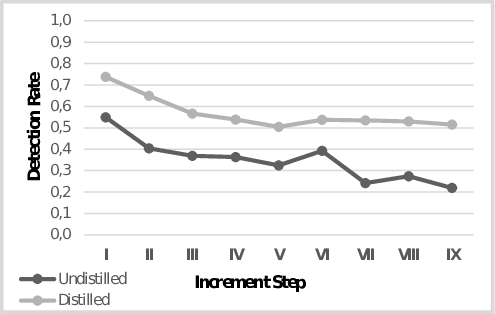}
            \caption{Group {\fontfamily{qcr}\selectfont 2a}: \textit{Duration \& Src\_Bytes}.}
            \label{fig:lines3}
        \end{subfigure}%
            \begin{subfigure}{\columnwidth}
            \centering
            \includegraphics[width=0.75\linewidth]{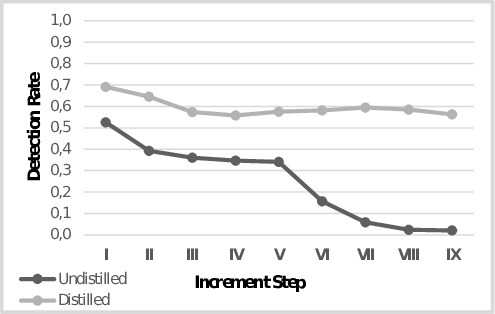}
            \caption{Group {\fontfamily{qcr}\selectfont 3a}: \textit{Duration \& Src\_Bytes \& Dst\_Bytes}.}
            \label{fig:lines4}
        \end{subfigure}
    
    \caption{Comparison of the detection rates on the adversarial datasets generated by all malware families.}
    \label{fig:comparison_full_lines}
\end{figure*}

\begin{figure*}[!htbp]
    \centering
    \begin{subfigure}{0.5\columnwidth}
        \centering
        \includegraphics[width=\linewidth]{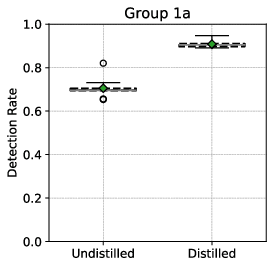}
        \caption{Boxplot for Fig.~\ref{fig:lines1}.}
        \label{fig:box1}
    \end{subfigure}\hfill%
    \begin{subfigure}{0.5\columnwidth}
        \centering
        \includegraphics[width=\linewidth]{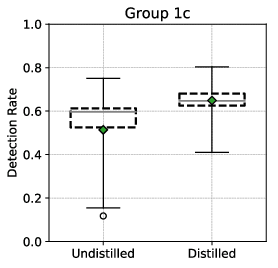}
        \caption{Boxplot for Fig.~\ref{fig:lines2}.}
        \label{fig:box2}
    \end{subfigure}\hfill%
    \begin{subfigure}{0.5\columnwidth}
        \centering
        \includegraphics[width=\linewidth]{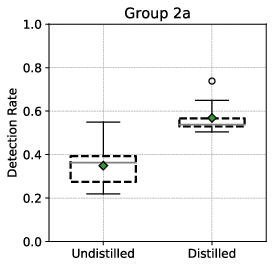}
        \caption{Boxplot for Fig.~\ref{fig:lines3}.}
        \label{fig:box3}
    \end{subfigure}\hfill%
        \begin{subfigure}{0.5\columnwidth}
        \centering
        \includegraphics[width=\linewidth]{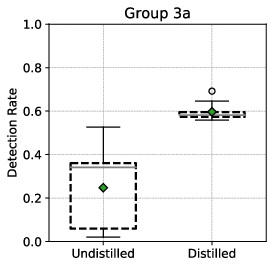}
        \caption{Boxplot for Fig.~\ref{fig:lines4}.}
        \label{fig:box4}
    \end{subfigure}
    
    \caption{Boxplot visualization of the results in Figs.~\ref{fig:comparison_full_lines}.}
    \label{fig:comparison_full_box}
\end{figure*}

We investigate the increased stability of our proposal through the fine grained comparisons in Figs.~\ref{fig:comparison_full_lines}, where the lines denote the detection rate (averaged for all botnet families) of the two models for four fixed groups of altered features (reported on top of each figure) and variable increment steps. The x-axis denotes the increment steps, and the y-axis the detection rate. The black and the gray line refers to the Undistilled and the Distilled model, respectively. In order to appreciate the improved stability of the performance, we include in Figs.~\ref{fig:comparison_full_box} the boxplots related to the results of Figs.~\ref{fig:comparison_full_lines}. These boxplots highlight that the Distilled detector is not affected by sudden performance drops, thus indicating that it is able to maintain its performance even against adversarial inputs that are different from the scenarios considered in this paper. 
The increased resilience of the Distilled detector is motivated by the fact that its Receiver model adopts a more robust set of feature importances when compared to the Undistilled model. In other words, a random forest model makes a prediction by comparing the features of a sample with the feature importances learned during its training phase: the probability labels used to train the Receiver produce a random forest model with a set of feature importances having a higher degree of flexibility than that of the Undistilled classifier, which adopts hard class labels. As a consequence, an adversary can significantly alter the detection results of the Undistilled model through tiny alterations of the features, while the Distilled detector is capable of withstanding even perturbations of high magnitude.
For example, let us consider two cases: in Fig.~\ref{fig:lines2} the adversary modifies only one feature (\textit{Dst\_Bytes}); in Fig.~\ref{fig:lines4} the adversary changes three features (\textit{Duration}, \textit{Src\_Bytes} and \textit{Dst\_Bytes}). In the former case, the two detectors have comparable performance for the first increment steps because the manipulated feature (\textit{Dst\_Bytes}) has high and similar importance for both models. In the latter instance, when alterations concern even incoming bytes and flow duration, the detection rates of the Undistilled model are unacceptably low (below 15\%).

\begin{figure*}[!htbp]
    \centering
    \begin{subfigure}{\columnwidth}
        \centering
        \includegraphics[width=0.75\linewidth]{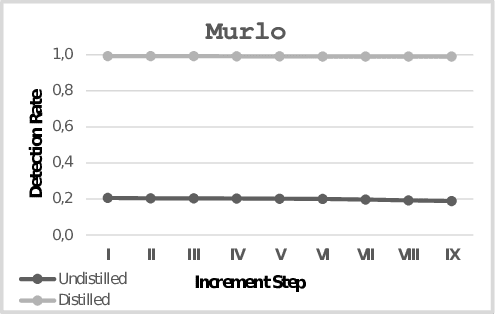}
        \caption{Group {\fontfamily{qcr}\selectfont 1a}: \textit{Duration}}
        \label{fig:1amurlo}
    \end{subfigure}%
    \begin{subfigure}{\columnwidth}
        \centering
        \includegraphics[width=0.75\linewidth]{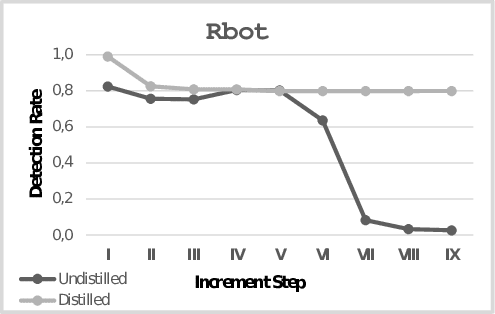}
        \caption{Group {\fontfamily{qcr}\selectfont 2b}: \textit{Duration \& Dst\_Bytes}}
        \label{fig:2brbot}
    \end{subfigure}
    
    \par\bigskip
    
    \begin{subfigure}{\columnwidth}
        \centering
        \includegraphics[width=0.75\linewidth]{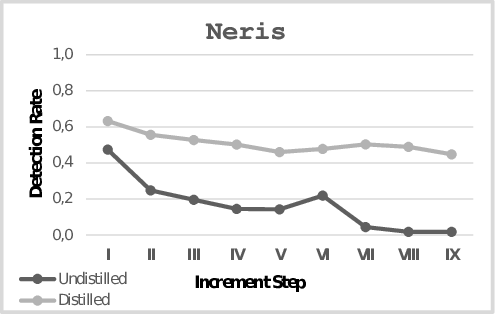}
        \caption{Group {\fontfamily{qcr}\selectfont 3a}: \textit{Duration \& Src\_Bytes \& Dst\_Bytes}}
        \label{fig:3aneris}
    \end{subfigure}%
        \begin{subfigure}{\columnwidth}
        \centering
        \includegraphics[width=0.75\linewidth]{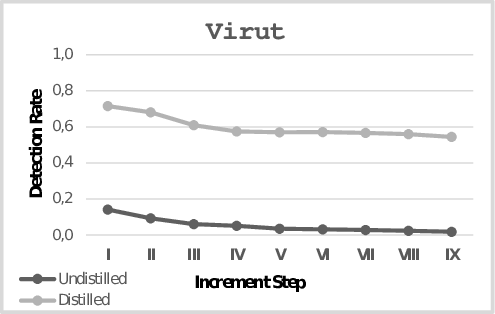}
        \caption{Group {\fontfamily{qcr}\selectfont 4a}: \textit{Duration \& Src\_Bytes \& Dst\_Bytes \& Tot\_Pkts}}
        \label{fig:4avirut}
    \end{subfigure}
    
    \caption{Comparison of the detection rates on the adversarial samples generated by specific malware families.}
    \label{fig:comparison_specific}
\end{figure*}

\begin{figure*}[!htbp]
    \centering
    \begin{subfigure}{0.5\columnwidth}
        \centering
        \includegraphics[width=\linewidth]{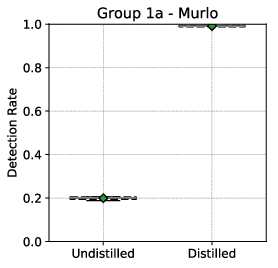}
        \caption{Boxplot for Fig.~\ref{fig:1amurlo}.}
        \label{fig:1amurlo_box}
    \end{subfigure}\hfill%
    \begin{subfigure}{0.5\columnwidth}
        \centering
        \includegraphics[width=\linewidth]{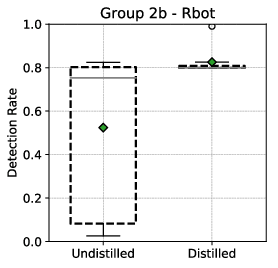}
        \caption{Boxplot for Fig.~\ref{fig:2brbot}.}
        \label{fig:2brbot_box}
    \end{subfigure}\hfill%
    \begin{subfigure}{0.5\columnwidth}
        \centering
        \includegraphics[width=\linewidth]{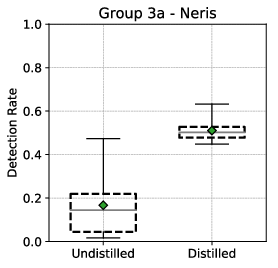}
        \caption{Boxplot for Fig.~\ref{fig:3aneris}.}
        \label{fig:3aneris_box}
    \end{subfigure}\hfill%
        \begin{subfigure}{0.5\columnwidth}
        \centering
        \includegraphics[width=\linewidth]{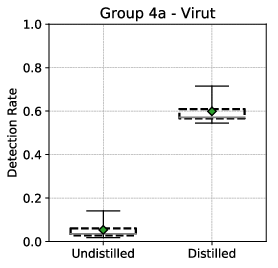}
        \caption{Boxplot for Fig.~\ref{fig:4avirut}.}
        \label{fig:4avirut_box}
    \end{subfigure}
    
    \caption{Boxplot visualization of the results in Figs.~\ref{fig:comparison_specific}.}
    \label{fig:comparison_specific_box}
\end{figure*}

The improved resilience of our method is confirmed by comparing the detection rates of the two detectors for fixed botnet families. The results and corresponding boxplots are presented in Fig.~\ref{fig:comparison_specific} and Fig.~\ref{fig:comparison_specific_box}, respectively. The name of the considered botnet family is reported on top of each figure. Overall, these figures confirm the superior detection capabilities and improved stability of the Distilled model.

\subsection{Comparison with existing defensive strategies}
\label{sec:comparison}

We compare the effectiveness of our proposal against two known countermeasures against evasion adversarial attacks that have been proposed in the literature~\cite{Apruzzese:Addressing, Anderson:DeepDGA, Zhang:Adversarial, Gardiner:Malware}, and that can be applied to any supervised machine learning algorithm: \textit{adversarial retraining} and \textit{feature removal}. To this purpose, we perform our experiments by following the same procedures described in~\cite{Apruzzese:Addressing}, due to the common characteristics shared by the considered adversarial scenarios and employed datasets. Hence, for the case of \textit{adversarial retraining} we generate a ``hardened'' Undistilled detector by re-training it after introducing a small ($10\%$) portion of the generated adversarial samples into the corresponding training sets, and then measure its detection rate on the same adversarial datasets used in our previous experiments for both the normal and adversarial scenarios. The results of this evaluation are presented in Table~\ref{tab:retraining} which shows the (averaged) Recall obtained by the re-trained Undistilled detector, the proposed Distilled detector, and the baseline Undistilled detector that we include for completeness. 

\begin{table}[!ht]
\centering
\caption{Comparison with adversarial retraining.}
\resizebox{0.7\columnwidth}{!}{
	\begin{tabular}{|c|c|c|}
		\hline
		\textbf{Detector Type} & \begin{tabular}{@{}c@{}}\textbf{Recall} \\ \textbf{(normal)}\end{tabular} & \begin{tabular}{@{}c@{}}\textbf{Recall} \\ \textbf{(adversarial)}\end{tabular} \\
		\hline\hline
		\cellcolor{gray!65}Undistilled (retrained) & \cellcolor{gray!65}$0.9695$ & \cellcolor{gray!65}$0.4987$ \\
		\cellcolor{gray!45}Undistilled (baseline) & \cellcolor{gray!45}$0.9684$ & \cellcolor{gray!45}$0.2573$ \\
		\cellcolor{gray!20}Distilled & \cellcolor{gray!20}$0.9751$ & \cellcolor{gray!20}$0.5152$ \\
		\hline
	\end{tabular}
}
\label{tab:retraining}
\end{table}

With regards to \textit{feature removal}, we develop a different Undistilled detector by training it on the same dataset used in our previous experiments but without considering the features that we modified to generate our adversarial samples (that is, \textit{Tot\_Pkts}, \textit{Duration}, \textit{Dst\_Bytes}, \textit{Src\_Bytes}), and then test it on the datasets used in Section~\ref{sec:normal}; this is motivated by the fact that \textit{feature removal} countermeasures, despite being resilient against adversarial attacks targeting the removed features, are known to generate excessive false alarms. The evaluation results are shown in Table~\ref{tab:removal}, which compares the (average) Precision, Recall and F1-score of the Undistilled detector (after excluding the features) with those obtained by the Distilled and the baseline Undistilled detector.

\begin{table}[!ht]
\centering
\caption{Comparison with feature removal.}
\resizebox{0.9\columnwidth}{!}{
	\begin{tabular}{|c|c|c|c|}
		\hline
		\textbf{Detector Type} & \textbf{F1-Score} & \textbf{Precision} & \textbf{Recall} \\
		\hline\hline
		\cellcolor{gray!65}Undistilled (feature removal) & \cellcolor{gray!65}$0.8728$ & \cellcolor{gray!65}$0.8497$ & \cellcolor{gray!65}$0.8974$ \\
		\cellcolor{gray!45}Undistilled (baseline) & \cellcolor{gray!45}$0.9729$ & \cellcolor{gray!45}$0.9774$ & \cellcolor{gray!45}$0.9684$ \\
		\cellcolor{gray!20}Distilled & \cellcolor{gray!20}$0.9777$ & \cellcolor{gray!20}$0.9804$ & \cellcolor{gray!20}$0.9751$ \\
		\hline
	\end{tabular}
}
\label{tab:removal}
\end{table}

By observing Table~\ref{tab:retraining}, we note that our proposal exhibits a higher detection rate in both scenarios. At the same time, concerning Table~\ref{tab:removal}, we appreciate that the Distilled detector achieves significantly better results. Indeed, we highlight that the proposed distillation method is not affected by the issues that characterize similar countermeasures: \textit{feature removal} strategies generate unacceptable rates of false positives, whereas \textit{adversarial retraining} requires to constantly update the training set with all the possible variations of samples that can be modified by the attacker (as explained in Section~\ref{sec:related}).

By taking into account all these analyses and evaluations, we can draw the following main conclusions.
\begin{itemize}
    \item Current state-of-the-art detection models based on machine learning have features that are too sensitive to the possibile manipulation of an attacker.
    \item The proposed variation of the defensive distillation technique can be used to devise random forest detectors that: achieve same or better detection performance than existing algorithms in scenarios that are not subject to adversarial inputs; exhibit improved robustness and stability against adversarial attacks; are not affected by the limitations of existing countermeasures.
    \item Although our proposal is an important result towards the reduction of the impact of adversarial inputs against machine learning detectors, it represents just a first step. There is still space for researches that aim to further improve the detection rates. 
\end{itemize}

\section{Conclusions}
\label{sec:conclusions}

Adversarial attacks represent a prominent and dangerous menace to organizations that rely on machine learning cyber detectors. We observe that existing approaches are based on classification criteria that are too rigid for the highly variable cyber security domain. The intuition is that by developing more flexible models it is possible to counter the manipulation of malicious samples. For this reason, we present an original method that limits the impact of adversarial perturbations by leveraging the defensive distillation technique. We consider the random forest algorithm due to its superior performance in cybersecurity detection tasks. An extensive campaign of experimental evaluations demonstrates the effectiveness of the proposed method, which achieves a twofold advantage over the state-of-the-art: in scenarios subject to adversarially manipulated inputs, it improves the detection rate up to $250\%$; in scenarios that are not subject to adversarial attacks, it achieves a similar or superior accuracy than existing techniques. This latter achievement is of particular importance because existing approaches that aim to counter adversarial attacks are often subject to a reduced performance in non-adversarial settings. Despite these promising results, our method presents room for further improvements. The proposed approach represents an original contribution to design robust detectors with high detection rates and strong enough against adversarial attacks.

\ifCLASSOPTIONcaptionsoff
  \newpage
\fi

\end{document}